\documentclass[a4paper,aps,pra,reprint,superscriptaddress, floatfix,twocolumn]{revtex4-2}
\usepackage[normalem]{ulem}
\usepackage[dvipsnames]{xcolor}
\usepackage{amsfonts}
\usepackage{physics}
\usepackage[english]{babel}
\usepackage{hyperref}  
\usepackage{mathtools}
\usepackage{listings}
\usepackage{bm}
\usepackage{latexsym}
\usepackage{times} 
\usepackage{graphicx} 
\graphicspath{{figures/}} 
\usepackage{booktabs} 
\usepackage{amsfonts, amsmath, amsthm, amssymb} 
\usepackage{graphicx}
\usepackage{bm}

\usepackage{soul} 

\begin{document}

\title{Qubit-environment entanglement outside of pure decoherence: hyperfine interaction}

\author{Tymoteusz Salamon}
\affiliation{ICFO - Institut de Ciencies Fotoniques, The Barcelona Institute of Science and Technology, Av. Carl Friedrich Gauss 3, 08860 Castelldefels (Barcelona), Spain}

\author{Marcin P\l{}odzie\'n}
\affiliation{ICFO - Institut de Ciencies Fotoniques, The Barcelona Institute of Science and Technology, Av. Carl Friedrich Gauss 3, 08860 Castelldefels (Barcelona), Spain}

\author{Maciej Lewenstein}
\affiliation{ICFO - Institut de Ciencies Fotoniques, The Barcelona Institute of Science and Technology, Av. Carl Friedrich Gauss 3, 08860 Castelldefels (Barcelona), Spain}
\affiliation{ICREA, Pg. Lluis Companys 23, Barcelona, Spain}

\author{Katarzyna Roszak}
\affiliation{Institute of Physics (FZU), Czech Academy of Sciences, Na Slovance 2, 182 21 Prague, Czech Republic}

\date{\today}

\begin{abstract}

In spin-based architectures of quantum devices, the hyperfine interaction between the electron spin qubit and
the nuclear spin environment remains one of the main sources of decoherence. This paper provides a short
review of the current advances in the theoretical description of the qubit decoherence dynamics. Next, we
study the qubit-environment entanglement using negativity as its measure. For an initial maximally mixed state
of the environment, we study negativity dynamics as a function of environment size, changing the numbers of
environmental nuclei and the total spin of the nuclei. Furthermore, we study the effect of the magnetic field on qubit-environment disentangling time scales.
 
\end{abstract}
\maketitle

\section{\label{sec:level1}Introduction}

Non-classical correlations, such as entanglement
are crucial resources for future quantum technologies \cite{Acin_2018,Eisert2020,Kinos2021,Laucht_2021,Zwiller2022,Fraxanet2022}. Entanglement and coherent superpositions of states of a given, well controlled quantum system are the key ingredients of future quantum computers. However, beside entanglement within the controlled quantum system, 
uncontrolled processes such as interactions with the environment which lead to decoherece, can also be a result of the formation of quantum correlations.
The effect of decoherence on the operation of a quantum algorithm is very detrimental, since
it leads to the quantum superposition behaving as a statistical mixture of states
\cite{Zurek2003, Schlosshauer2005}. The result is that the gain from
the use of quantum resources during computation is diminished. 
As such, finding methods to minimize decoherence and restore quantumness is very important
for applicative quantum technologies.

One of the critical directions in modern quantum sciences is research on entanglement generation between the controlled quantum system and its environment in order to understand how to overcome limitations of quantum devices due to decoherence processes.
Understanding decoherence lies at the heart of measurement, quantum information processing, and, more fundamentally, the transition from the quantum to the classical world. One of the important aspects of system-environmnet entanglement generation is a situation when
the entanglement is generated during the joint qubit-environment evolution
with the sole exception of the case when the initial state reduces the evolution to pure
decoherence. There is as qualitative change between a small environment regime and the
situation when its Hilbert space is larger either by means of many nuclei or large nuclear
spin. The exception here is a single nuclear spin environment with a large spin, which
displays much more structured behaviors. The application of the magnetic field,
which significantly changes the free Hamiltonian of the qubit, has a strong impact on the
evolution of entanglement, leading to much more complex time-dependencies. 

In this paper we present current status in the field of entanglement generation between controlled spin-qubit and its environment, and we present qualitative predictions on disentanglement times as a function of environment size.

The paper is structured as follows. In Sec.~ \ref{sec:level2} we present a brief review of achievements in the field of decoherence processes between quantum systems and their environments. Next, in Sec.~ \ref{sec:System} we present the simplest
 Hamiltonian of a spin-qubit interacting with a spin-bath.
In Sec.~\ref{sec:entanglement_measures} we concisely introduce the entanglement
measure used for qubit-bath entagnlement characterization. In Sec.~\ref{sec:Results} we present the results, analizing the 
evolution of Negativity for different parameters of the environment, initial states of 
the qubit, and values of applied magnetic field. 
Finally, we conclude in Sec.~\ref{sec:Conclusions}.
\section{\label{sec:level2}State of the art}

  The simplest system-environment models consist of a central spin  or qubit  coupled to the environment modeled by an ensemble of $1/2$-spins denoted as a spin-bath \cite{Prokof2000,Hutton2004,Breuer2004,Hamdouni2006,Yao2007,Yao2006,Yao2008,Cywinski2009,Bortz2010,Stanek2013,Bragar2015}. This model is a very good theoretical platform allowing studies on system-environment entanglement for one of the most promising candidates for quantum computation, solid-state spin systems \cite{Loss1998,Immamoglu1999,Culcer2010}, which are inevitably coupled to their surrounding environment, usually through interactions with neighboring nuclear spins and are vulnerable to decoherence processes. Over the last years much research was focused on spin-bath entanglement  \cite{mazurek14a,mazurek14b,Krzywda2018,strzalka20}, qubit decoherence reduction during dynamical decoupling \cite{Khodjasteh2005,Viola1998,Viola2005,Uhrig2007,Lee2008,Yang2008,Du2009,Cywinski2014,Szankowski2017,Szankowski2018,Krzywda2019,Sakuldee2020}, decoherence supression due to spin coupilng to the self-interacting bath \cite{Tessieri2003,Schliemann2003,Olsen2007,Lai2008}, entanglement sharing \cite{Dawson2005}, coupilng to decoherence free subspaces \cite{Duan1997,Zanardi1997,Lidar1998}, and disentangelment  \cite{Wi2014}.

For pure states of both the qubits and the environment, decoherence can only be a 
result of entanglement being formed \cite{Zurek2003}. As the states of most environments are mixed,
there is a distinction between decoherence resulting from classical noise
\cite{helm09,helm10,lofranco12,hilt09,pernice11,pernice12}
and the more quantum processes, for which decoherence is a direct consequence of the 
build-up of quantum correlations \cite{eisert02,hilt09,pernice11,pernice12,maziero12,costa16,salamon17,strzalka20}. Both types are abundant in nature and for the most part, decoherence processes have never
been qualified in this context. This is because the study of the amount
of quantum correlations in systems which are large and mixed is very taxing numerically.

Considering the recent progress in the control and operation of quantum systems, the distinction
between classical and entangling decoherence processes becomes important. This is because 
the presence of entanglement with the environment is likely to lead to measurable changes in the evolution
of a system of qubits which is interspaced by gate operation and measurements, in
the same way, as desired quantum correlations between the qubits can lead to the increased
capabilities of quantum devices. Uncontrolled entanglement can in turn
lead to unpredicted errors in the operation of quantum algorithms.

In the case of pure decoherence, quantification of entanglement with the environment
can be grossly simplified \cite{roszak15,roszak18,roszak20}. This allowed for entanglement
to be unambiguously connected with the transfer of information about the state of the qubits
into the environment and consequently, for methods for the direct detection of such
entanglement to be proposed \cite{roszak19a,rzepkowski20,strzalka21,roszak21}
and entanglement to be measured \cite{zhan21}.
These methods involve basic operations and measurements performed on the qubit subsystem,
and rely on the fact that decoherence which is the result of entanglement generation
alters the state of the environment in a distinctly different manner than classical noise. 
Since entanglement in pure decoherence can now be studied on a very general level
\cite{roszak15,roszak18}, the effect of system-environment entanglement on quantum teleportation
has been studied \cite{harlender22}, yielding an effect when decoherence
can counter-intuitively be used for purification of the qubit state \cite{roszak22}.

In this paper, we will focus on the study of quantum correlations between a qubit
defined on a localized electron spin interacting with an environment of nuclear spins via the hyperfine interaction. Such a coupling does not lead to pure decoherence unless the magnetic field yields a very large splitting between the qubit states or the initial qubit state is an equal superposition state,
while it is a very fundamental decoherence process for solid state spin qubits.
It can be used to describe decoherence for
charge carrier spins confined in quantum dots \cite{bechtold16,nichol17,yoneda18,watzinger18,hendrickx20}, 
NV-center spin qubits \cite{togan10,degen17,wood18,tchebotareva19,wang20b},
and spin qubits on donor atoms \cite{pla12,miao20,madzik21,fricke21}.

Outside of pure decoherence the means for the study of quantum correlations
with an environment remain highly
numerical regardless of the chosen entanglement measure. This means that it is necessary
to study a concrete ensemble and set of initial states, as general the behavior of a given
form of interaction remains out of reach. We will study quantum correlations
generated between the spin of an electron confined in a GaAs quantum dot
interacting with an environment composed of $F=3/2$ nuclear spins of the gallium and arsenite atoms
\cite{merkulov02,barnes11,urbaszek13}, but as our aim is the more general study 
of entanglement during this type of decoherence processes, we will use the nuclear
spin as a parameter.
We focus on the situation when the nuclear bath is initially at an infinite temperature
equilibrium. This case is realistic for such qubits assuming that the environment
has not been especially prepared, since the nuclear Zeeman terms for both types of atoms
are very small compared to the thermal energy at typical experimental temperatures
\cite{barnes11,cywinski11}. It is also an interesting limit from the point of view
of entanglement theory, since the environment is as classical as possible. In case of
pure decoherence, no quantum correlations with a qubit 
can be formed if the environment is fully mixed
\cite{roszak15,roszak17}, but entanglement is possible for a qutrit and larger systems \cite{roszak18}. The limitation on qubit-environment entanglement for infinite 
temperature environments does not hold outside of pure-decoherence
(an example with an ``environment'' composed of a single qubit can be found 
in the appendix of Ref.~\cite{roszak15}).
\section{\label{sec:System}The qubit and the environment}

The qubit-environment system under study consists of an electron confined in a lateral
GaAs quantum dot. The qubit is defined on the spin states of the electron and is initially
in a pure state. The environment consists of nuclear spins of the surrounding atoms,
initially in a maximally mixed state due to the very small Zeeman splitting characteristic
for such nuclear spins. They interact via the hyperfine coupling, which is the dominant
spin-spin coupling here.

Let us start with a general Hamiltonian for the single qubit interacting with the spin-bath, i.e.
\begin{equation}\label{eq:H_general}
    \hat{H} = \hat{H}_S + \hat{H}_B + \hat{H}_{BB} + \hat{H}_{SB},
\end{equation}
where $\hat{H}_S$ is the Hamiltonian for a central spin, $\hat{H}_B$ is a Hamiltonian for the spin-bath (also denoted as environment) consisting of many spins, $\hat{H}_{BB}$ describes interaction between spins in the bath, and $\hat{H}_{SB}$ is the qubit-bath coupling term. The qubit spin operator $\hat{\mathbf{S}} = (\hat{S}_x,\hat{S}_y,\hat{S}_z)$ is defined on an electron confined in a quantum dot (QD) interacting with
an ensemble of environmental nucleus via the hyperfine interaction, where $k$-th nuclei is decribed by the spin operator $\hat{\mathbf{F}}_k = (\hat{F}_{x,k},\hat{F}_{y,k},\hat{F}_{z,k})$. The respective parts of the total Hamiltonian \eqref{eq:H_general} read
\begin{equation}
\label{main_hamiltonian}
\begin{split}
    \hat{H}_S & = -g\mu_B {\cal B}\hat{S}_z,\\
    \hat{H}_B & = E_I\sum_{k=1}^{n} \hat{F}_{z,k},\\
    \hat{H}_{BB} & = \sum_{k\ne l}^n d_{kl}\hat{\mathbf{F}}_k\hat{\mathbf{F}}_l,\\
    \hat{H}_{SB} & = \sum_{k=1}^n A_k \hat{\mathbf{S}}\hat{\mathbf{F}}_k.
    \end{split}
\end{equation} 
 
The $\hat{H}_S$ term is the Zeeman splitting of the central spin (qubit)
under the assumption that the magnetic field ${\cal B}$ was applied in the $z$ direction
(perpendicular to the plane of the QD). Here, $\mu_B$ is the Bohr magneton,
and $g=-0.44$ is the electron g-factor.  The Zeeman term for the environment, $\hat{H}_B$,  can be omitted due to the small magnitudes
of the splitting per Tesla of magnetic field
compared to the thermal energy at temperatures characteristic for
experiments on QD spin qubits \cite{Abragam1961,merkulov02,Bluhm11}.
The  $\hat{H}_{BB}$ term describing dipolar interaction between  nuclear spins in the system under study is also known to be negligible and has been omitted \cite{merkulov02}.  The $\hat{H}_{SB}$ term is the hyperfine interaction with the spins of the environment.
Finally, the Hamiltonian describing the qubit and environment is given by
\begin{equation}
\label{eq:H_general}
    \hat{H} = \hat{H}_S + \hat{H}_{SB} = -g\mu_B {\cal B}\hat{S}_z + \sum_{k=1}^n A_k \hat{\mathbf{S}}\hat{\mathbf{F}}_k,
\end{equation}
where the coupling constants $A_k$ describe the contact hyperfine interaction between the electronic spin and $k$-th nucleus. In general they are defined as $A_k = A_k^0 V_0\abs{\psi(\textbf{r}_k)}^2$, with $A_k^0=\frac{2}{3}\gamma_e\gamma_k\mu_0$.
Here, $V_0$ is the unit cell volume, $\psi(\textbf{r}_k)$ is a value of the electronic wave function at the position of the k-th nucleus and $\gamma_{k/e}$ denote the nuclear and electronic giro-magnetic ratios, respectively \cite{Liu_2007}. 
We use parameters characteristic for lateral GaAs QDs which
are presented in Table \ref{table}.
The envelope of the electron wave function $\psi(\textbf{r}_k)$
is modeled by an anisotropic Gaussian reflecting the shape of the QD, with in-plane width
$l_{\perp}=20$ nm and $l=2$ nm in the $z$-direction. 

\begin{table}[h]
 \begin{center}
\begin{tabular}{cccc}
& $Ga^{69}$  & $Ga^{71}$ & $As^{75}$\\
\hline \\[-.5em]
Abundance  & \hspace{.005cm} $60.4\%$ \hspace{.005cm} & \hspace{.005cm} 59.6\% \hspace{.005cm} & \hspace{.005cm} 100\% \hspace{.005cm} \\
\hline \\[-.5em]
Spin moment $F$ & $3/2$ & $3/2$ & $3/2$ \\
\hline \\[-.5em]
$A_\alpha^0 [ \mu \mathrm{eV}]$ & $36 $ & $46$ & $43$ \\
\hline \\[-.5em]
$\gamma_k [10^{7} \mathrm{rad T^{-1} s{-1}}]$ & $6.44 $ & $8.18$ & $7.29$ \\
\hline \\[-.5em]
\multicolumn{4}{c}{$\gamma_e=0.176 \times 10^7 [\mathrm{rad T^{-1} s^{-1}}] $}\\
\multicolumn{4} {c}{$V_0=800 [\mathrm{nm}^3]$}
\end{tabular}
\end{center}
\vspace{-.25cm}\hspace{.3cm}
\label{Table}
\caption{Material parameters of the atomic components of GaAs quantum dots used to calculate the coupling constants between the nuclear bath and the spin qubit.\label{table}}
\end{table}
In the following we will use the common approximation used for Hamiltonian (\ref{eq:H_general}),
namely the box model \cite{khaetskii03,zhang06,mazurek14b,chesi15}.
This entails replacing all coupling constants $A_k$ in the interaction by their average value,
which for the QD under study yields
$A_k=A \approx 83 \mu eV$ \cite{mazurek14b}.
The second term in Hamiltonian (\ref{eq:H_general}) can now be written as 
\begin{equation}
    \hat{H}_{SB} \approx A \hat{\mathbf{S}}\hat{\mathbf{K}},
\end{equation}
where $\hat{\mathbf{K}}$
denotes the total bath spin operator, 
$\hat{\mathbf{K}}=\sum_{k=1}^{n} \hat{\mathbf{F}}_k$.
The approximation allows the Hamiltonian to be diagonalized analytically,
since it can be represented in a block-diagonal $2\times 2$ matrix form.

Let us denote the qubit states as $\ket{\uparrow}$ and $\ket{\downarrow}$
and the environmental basis states, which are the eigenstates of both the
total bath spin operator and its $z$ component, $\ket{K,m}$. Using these bases
we can 
write the eigenstates of the Hamiltonian (\ref{eq:H_general}) in the box model approximation
as 
\begin{subequations}
	\label{eq.5}
\begin{eqnarray}
   \ket{+Km} &=& \cos\theta_{K,m}\ket{\uparrow K m}+\sin\theta_{K,m}\ket{\downarrow K m+1}, \\ 
   \ket{-Km} &=& -\sin\theta_{K,m}\ket{\uparrow K m}+\cos\theta_{K,m}\ket{\downarrow K m+1},
\end{eqnarray}
\end{subequations}
with
\begin{equation}
\sin \theta_{K,m} = \frac{M_{K,m}}{\sqrt{(E^+_{m}+E_{m+1})^2+M_{K,m}^2}}, 
\end{equation}
where the elements within each block of the matrix are given by
\begin{align}
   & E_m = \frac{\hbar}{2}(-g\mu_B {\cal B}/h+A m),\nonumber\\
   & M_{K,m} = \hbar A/2\sqrt{K(K+1)-m(m+1)},
\end{align} 
and the eigenvalues corresponding to eigenvectors (\ref{eq.5}) have the form
 \begin{equation}
   E_{K,m}^{\pm} = \frac{1}{2}\Big{(}-\hbar A/2 \pm \sqrt{(E_m+E_{m+1})^2+4M_{K,m}^2}\Big{)}  .
 \end{equation}

Since the eigenvectors and eigenvalues of the Hamiltonian are all known, regardless of the
size of the environment, the evolution of any initial qubit-environment state can 
be found at any time. 
In the following, we will always be studying a product initial state, where the state of
the qubit is given by
\begin{equation}\label{eq:psi_initial}
\ket{\psi}=\alpha\ket{\uparrow}+\beta\ket{\downarrow}
\end{equation}
and the density matrix of the environment is a product of nuclear spin states, in which
the state of each nucleus is a maximally mixed state (infinite-temperature Gibbs state \cite{Nitzan2006}). 

Transforming the state of the environment into the total nuclear spin basis $\ket{K,m}$
has to take into account the multiple ocurrences of certain states due to the rules
of addition for angular momentum, so the density matrix can be written as
\begin{equation}
\label{eqn:rho_E_initial}
\hat{\rho}_E(0) = \sum_{m} P_{K,m}|Km\rangle\langle Km|,
\end{equation}
with probabilities given by \cite{Gottlieb77,Mik77}
\begin{equation}
\label{prob}
P_{K,m} = \frac{1}{{\cal Z}} \sum_k(-1)^k\binom{n}{k}\binom{mn+n-2mk+k-K-2}{n-2},
\end{equation}
where ${\cal Z}$ is the normalization constant.

\section{Entanglement measures - Negativity}\label{sec:entanglement_measures}

Since the system is bipartite and  built of qubit and environment having in principle many degrees of freedom we use the Negativty ${\cal N}$ \cite{vidal02,Plenio_2005} as an entanglement measure, which is based on Positive Partial Transpose (PPT) criterion \cite{peres_96,horodecki96}. However, one cannot exclude a possibility of a bound entanglement formation \cite{Horodeccy_98}. This in principle could be problematic, since Negativity ${\cal N}$ is blind to this kind of entanglement, but since the qubit is always initially in a pure state
while the environment is maximally mixed, the purity of the whole system throughout the
evolution is twice the minimum purity for a given system size, and it has been shown
in Ref.~\cite{kraus00} that bound entangled states cannot occur in this case.

Based on Peres-Horodecki criterion of separability, Negativity is defined as the sum of the negative eigenvalues of the partially transposed density matrix. It is irrelevant with respect to which subsystem the partial transposition is being made. Negativity can be written in the general form
\begin{equation}\label{eq:negativity}
    {\cal N}(\rho) = \sum_i\frac{\abs{\lambda_i}-\lambda_i}{2},
\end{equation}
where $\rho$ is the density matrix of the system and $\lambda_i$ are eigenvalues of $\rho^{\Gamma_A}$ where $\Gamma_A$ denotes partial transposition with respect to subsystem $A$ (which in our case is the qubit). The structure of the density matrix obtained during the evolution allows for simple partial transposition with respect to the qubit subspace by flipping two off diagonal quarters of the total density matrix.

\section{\label{sec:Results}Results}

In the following we study the evolution of qubit-environment entanglement as a function
of different factors, such as the magnetic field, which impacts the splitting between
the energy levels of the qubit. The initial state of the qubit is also a factor,
because it influences the nature of the evolution, and for an equal superposition state,
the hyperfine interaction only leads to pure decoherence. 
A relevant factor is also the number of
nuclear spins $n$. Since the calculation of negativity requires diagonalization of matrices
of the same dimension as the qubit-environment density matrix, there is a limit 
on the size of the environment that can be studied. Nevertheless, it is possible to 
model the decoherence of large environments with the help of a manageable number of nuclei.
Since the study of entanglement is much more dependent on the specific parameters of
the qubit-environment density matrix, it is a reasonable expectation that the convergence
of results to the infinite-reservoir limit will be much slower than in case of decoherence
and may not emerge at all. 

Another parameter is the spin of each nucleus $F$;
this changes the size of the single nucleus density matrix and modifies the nature of
the interaction with the qubit (allowing different transitions within the subspace of
each nucleus and affecting the coupling strengths). Although in GaAs, all nuclei
have spin $F=3/2$, we are interested in more general qualities of entanglement evolution
driven by the hyperfine interaction and change this parameter freely. It is relevant to note that the same decoherence
can be modeled by environments with different total spin of each nucleus,
but the evolution of entanglement shows qualitative differences for different species
of nuclei. 
\\
\\
Negativity ${\cal N}(\hat{\rho})$ is bounded from above by the negativity of the isotropic state 
\begin{equation}
    \hat{\rho}_{iso} = \kappa\dyad{\Phi^{+}}{\Phi^{+}}+(1-\kappa)\frac{1}{N}\mathbf{1},
\end{equation}
parametrized by a scalar $\kappa$, where $\ket{\Phi^{+}}$ is a maximally entangled state, denoted as ${\cal N}_{iso}$. Keeping in mind, that for any two-qubit state $\hat{\rho}$ for which $\tr[\hat{\rho}^2] = \tr[\hat{\rho}_{iso}^2]$, the following holds
${\cal N}_{\text{iso}} \ge {\cal N}(\hat{\rho})$ \cite{Horodecki99}. Next, noticing that the purity of the smallest system considered here is $\tr[\hat{\rho}^2]=1/2$, which  equals to the purity of the isotropic state with $\kappa \sim 0.5774$ and finally, noticing that entanglement between the qubit and the large spin-bath cannot be larger than between two qubits, the maximal negativity in the considered qubit-bath system is given by ${\cal N}_{iso} \approx 0.3661$.
 
Besides the evolution of the negativity ${\cal N}$ we focus on time scale $\tau$ on witch qubit-bath disentanglement appear, and its scaling with the total bath spin number $K = nF$.

\subsection{Single nucleus bath}

\begin{figure}[h!]
\includegraphics[width=1\linewidth]{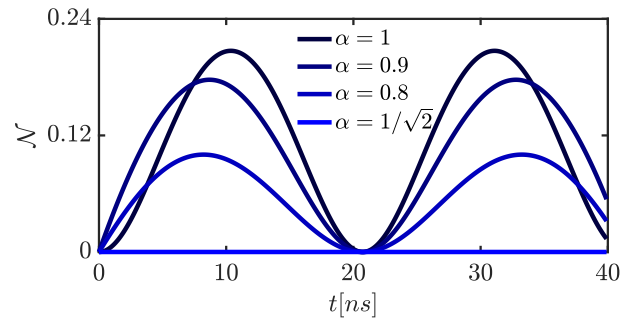}
\caption{
Negativity dynamics for a single nucleus bath with $F=1/2$ and four different initial states of the spin-qubit with $\alpha = \{1, 0.9, 0.8, 1\sqrt{2}\}$, eq.~\eqref{eq:psi_initial}. Maximal value of the negativity corresponds to $\alpha=1$ and vanishes for equal superposition $\alpha=\beta=1/\sqrt{2}$; in such case the system is separable during the whole evolution. The negativity period remains constant for each $\alpha\ne 1/\sqrt{2}$. 
\label{fig:fig1}}
\end{figure} 

Let us start with the situation when the environment consists of only one nuclear spin
with $F=1/2$. This means that we are dealing with a two-qubit scenario, in which
one qubit is initially in a pure state, while the other is maximally mixed, so the
initial purity of the whole state is half of the maximum two-qubit purity. 
 The negativity ${\cal N}$ has a time-periodic structure and vanishes  with period $\tau$, where
\begin{equation}
    \tau \approx \frac{3}{\hbar\sqrt{A^2 + g^2 \mu_B^2 {\cal B}^2}},
\end{equation}
 indicating disentanglement of the spin-qubit from its single spin $F=1/2$ bath. 

Fig.~\ref{fig:fig1} presents negativity evolution at zero magnetic field  for different initial states of the central qubit, eq.~\eqref{eq:psi_initial}. The interaction
is symmetric with respect to the spin-flip of the qubit, so only parameters $\alpha$ between
$\frac{1}{\sqrt{2}}$ and $1$ are of interest. Maximal value of the negativity corresponds to $\alpha=1$, while entanglement is never generated for the equal superposition initial qubit
state $\alpha=\beta=1/\sqrt{2}$; in such case the system is separable during the whole evolution. In these two cases, negativity can be found exactly. For $\alpha=1$ this is
because the two-qubit system is an X-state throughout the evolution
and entanglement is generated through transitions into the $\ket{\uparrow\downarrow},
\ket{\downarrow,\uparrow}$ subspace. For $\alpha=1/\sqrt{2}$ the hyperfine interaction
leads to fundamentally different processes and leads to pure decoherence for the 
central spin. The evolution is separable because it is not possible for entanglement
to be generated, if the environment is initially in a maximally mixed state
regardless of environment size \cite{roszak15}. For intermediate values of $\alpha$,
the two-qubit density matrix is not sparse enough for analytical calculation of negativity,
but the interaction does not lead to pure decoherence and transitions between different
two-spin states which lead to entanglement generation do occur at a lesser degree 
than for the spin-up  or spin-down states.

\begin{figure}[h!]
	\includegraphics[width=1\linewidth]{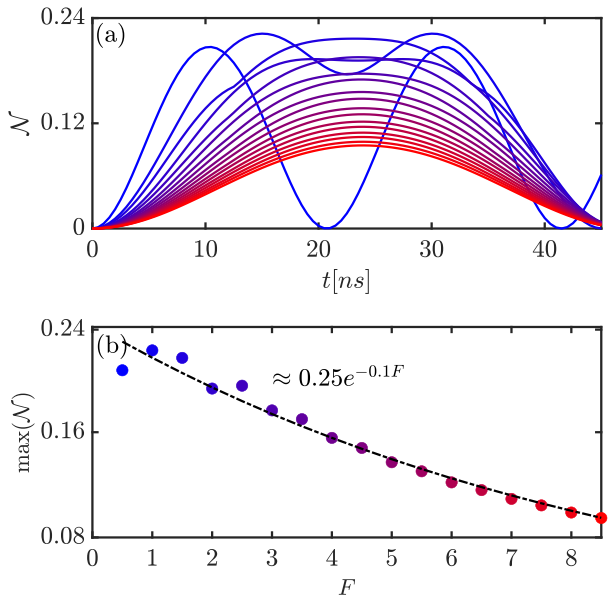}
	\caption{Negativity dynamics of the spin-qubit in state $\ket{\uparrow}$ and the bath consisting of a single nucleus with spin $F=1/2$ (blue) to $F=17/2$ (red), with a step of $1/2$. Panel (b) shows decay of the maximum value of the Negativity and a corresponding fit showing exponential decay for higher values of $F$ (black, dashed line).}
	\label{fig:fig2}
\end{figure}
In Fig.~\ref{fig:fig2}(a) we show the same evolution of entanglement while varying the 
total spin of the single environmental nucleus, $F$. The initial state of the qubit is always
$\ket{\uparrow}$, yielding the maximum possible negativity of all initial states. 
The nuclear spin varies from $F=1/2$ to $F=17/2$ with a step of $1/2$. Interestingly,
there is a large qualitative change in the behavior of entanglement when the spin is small
(between $F=1/2$ and $F=5/2$), but for larger spins the only effect of the increased 
size of the Hilbert space of the environment, is that negativity is smaller, while
growth and decay are qualitatively the same and occur at the same time-intervals
for $F \ge 5/2$. This is reflected well on Fig.~\ref{fig:fig2}(b)
where maximum negativity is plotted as a function of $F$. At larger values,
the decay is well fitted by an exponential function. 

\subsection{Bath of many nuclei}

We further study the dependence of entanglement evolution as a function of the growing
Hilbert space of the environment, which is taken into account in two distinct ways.
Firstly, the total spin number $F$ is varied as before, but also the number of nuclei $n$
is increased.
\begin{figure}[h!]
	\includegraphics[width=1\linewidth]{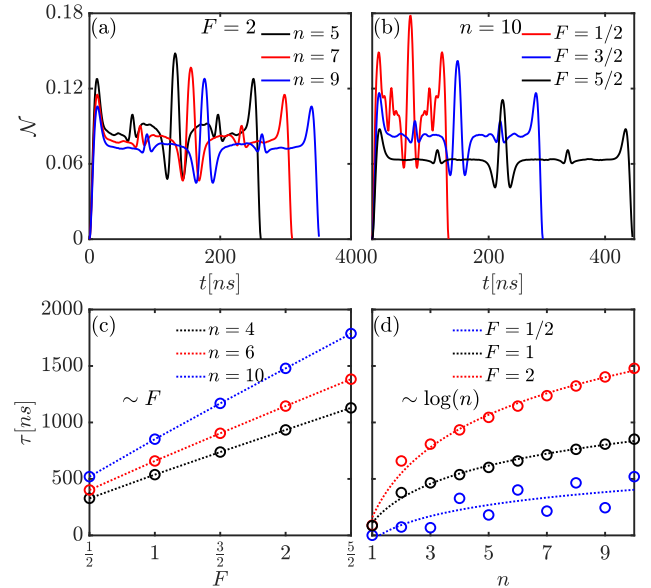}
	\caption{
		 Top row: Negativity dynamics for varying number of environmental spins for $F=2$, panel (a), and different total
		spin numbers $F$ for a constant number of nuclei $n=10$, panel (b), at zero magnetic field. Both panels only show one period of the evolution, which would restart after
		zero entanglement is reached.
	    Bottom row, panels (c,d): dots show time $\tau$ at which a separable state is
		obtained. Panel (c) presents $\tau$ as a function of 
		  $F$ for different values of $n$, with linear scaling $\tau \propto F$ (dotted lines). Panel (d) presents $\tau$ as function of number of nuclear spins $n$ for given $F$, with scaling $\tau \propto \log(n)$ (dotted lines). }
	\label{fig:fig3}
\end{figure}
In Fig.~\ref{fig:fig3} (a), negativity evolution is plotted for 
a fixed value of the spin of each nucleus $F=2$ for different values of the
number of environmental spins, $n=5,7,9$, while in Fig.~\ref{fig:fig3} (b),
the number of spins is constant $n=10$, while the spin of each nucleus is varied,
$F=1/2,3/2,5/2$. 
One can observe that the evolution
is qualitatively different from the situation when there is only one nucleus of the
environment, which is to be expected, since the nature of the initial state changes
drastically when a second nucleus is added. This is because of the rules for addition
of angular momenta in quantum mechanics, which result in multiple occurrences of the 
same total angular momentum $\ket{K,m}$ states, which is reflected in the probabilities
(\ref{prob}), while for a single spin, $K=F$ and the probabilities for each state of the
environment are equal in the infinite-temperature Gibbs state.
 
What is more surprising is that when the Hilbert space of the environment is large (outside of
$n=1$), there is no qualitative difference in the evolution for different
parameter choices. A bigger environment
(obtained by either a larger $n$ or $F$) results in a longer time until the whole
qubit-environment system returns to its initial state
(and then the evolution continues, which is not plotted in the figures for the sake of 
clarity) and a smaller maximum negativity which is reached, but the curves can be easily and with a reasonable degree of accuracy
transformed into one another by rescaling of the time and varying the amplitude.  Fig.\ref{fig:fig3}(c,d) present time $\tau$ at which a separable qubit-bath state is obtained as a function of the bath size. Fig.\ref{fig:fig3}(c) presents $\tau$ as a function of bath spin number $F = 1/2,1,3/2,2,5/2$ for fixed total nuclei in bath $n = 4,6,10$, which has linear scaling $\tau \propto F$. Panel Fig.\ref{fig:fig3}(d) shows $\tau$ as a function of $n$ for fixed $F = 1/2, 1,2$ which scales as $\tau\propto\log(n)$.
 For the total spin of the bath, $K = nF$ with $F = 1/2$, the disentanglement time scale $\tau$ depends on fermionic/bosonic nature of the bath. In particular, the bosonic bath (even number $n$) manifest larger entangling times than  fermionic counterparts (odd $n$). Such a behvaior is related to existence of the zero $z$-component of the spin, i.e. element $m=0$, in the absence of the magnetic field
\begin{equation}
    \sin\theta_{K,0} =\frac{\sqrt{K}}{\sqrt{2K+1}},
\end{equation}
which contributes to the evolution of the off-diagonal density matrix elements. For large bath spins, and high $K$, $\sin\theta_{K,0}\approx1/\sqrt{2}$ leading to equal superposition in this spin sector and as a consequence - longer entangling times. 
\subsection{\label{magnetic_field} External magnetic field}
\begin{figure}
    \centering
    \includegraphics[scale=0.35]{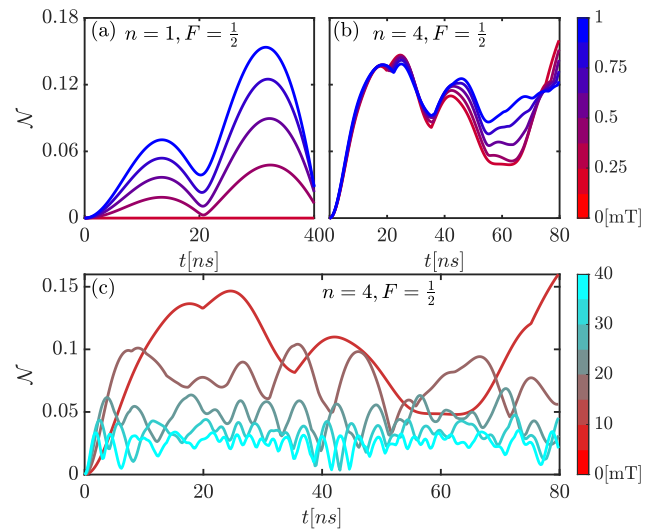}
\caption{
Evolution of the negativity between the qubit in initially equal superposition and the bath consisted of $n = 1$, panel (a), or $n = 4$, panels (b,c) nuclei with $F=1/2$ under different values of magnetic field $\cal{B}$. As one can observe, birth of entanglement on panel (a) occurs due to non-zero external magnetic $\cal{B}$. In the regime of relatively small magnetic fields, entanglement grows, as exemplified on the panels (a) and (b). However, such behavior is non-monotonic and in the regime of strong magnetic fields the system becomes practically uncoupled due to high energy gap between the magnetic sublevels. This dependence is plotted on panel (c), where the red ($\cal{B}=0$) line the same red line plotted on panel (b). One can immediately observe that even though the entanglement initially grows with increasing magnetic field, it starts to decay when $\cal{B}$ reaches significant magnitudes.}
\label{fig:fig4}
\end{figure}

Applying a finite magnetic field causes a splitting of the qubit eigenstates
and leads to a qualitative difference in the evolution of entanglement
already for the two-qubit case (when the environment is a single nucleus with spin $1/2$).
This is plotted in Fig.~\ref{fig:fig4} (a), where we observe a departure from sine 
behavior and a larger value of negativity at mid-range times, proportional to the
value of the magnetic field. 
Much more complex evolutions are observed in the presence of the magnetic field
for moderately larger Hilbert spaces of the environment, as plotted in Figs
\ref{fig:fig4} (b,c) for nuclear spin $1/2$ and $n=4$ nuclei,
and relatively small (a) and strong (c)  magnetic fields.
These plots show distinct points of 
non-differentiability 
which are characteristic to evolutions of negativity. Applying the magnetic field
yields curves which are much less structured and resemble the curves charateristic 
for the bigger environments more.

\section{Conclusions}\label{sec:Conclusions}

We have studied the generation of entanglement in a qubit-environment setup
where the interaction is driven by the hyperfine coupling, which
under most circumstances does not lead to pure decoherence. The initial state of the 
qubit is pure while the environment is maximally mixed. 
We have shown that when the decoherence does involve transitions between the qubit states,
it is accompanied by the creation of entanglement, even though the environment
is as classical as possible for its size. This cannot be the case for pure decoherence
and when the evolution reduces to pure decoherence (for an equal superposition state
of the qubit), we show that entanglement was not generated.

We used material and structure parameters characteristic for lateral GaAs quantum dots
to calculate the strength of the interaction and worked in the box model approximation
which is known to reproduce hyperfine-induced decoherence well. On the other hand,
we treated the number of environmental spins as well as the total spin of each nucleus
as a parameter, in order to be able to describe and understand transitions from
small to (relatively) large environments. We have observed that the case of a single
nuclear spin is special in the sense that increasing the Hilbert space of the environment
by enlarging the total spin of a single nucleus does not lead to the emergence
of more complicated entanglement dynamics characteristic for big environments
an above some threshold spin, there is not qualitative difference connected
with growing total nuclear spin. Otherwise, these behaviors emerge similarly
regardless if the Hilbert space is increased by the increase of the total spin of a 
given number of particles, or by increasing the number of nuclei. 
It is relevant to note, that a bigger environment leads to less entanglement
and slower evolution, in accordance to expectations.

Applying a finite magnetic field, which makes the free Hamiltonian of the qubit nontrivial,
results in much more complicated evolutions of entanglement regardless of all other parameters.
Even for the effectively two-qubit case, there is a non-trivial effect resulting from the
splitting of the qubit states, and the larger the environment, the more complex features are
displayed.  

We have shown that the evolution of qubit-environment 
entanglement driven by the hyperfine Hamiltonian yields qualitatively different
results than are possible due to any Hamiltonian that leads to pure decoherence.
In opposition to pure decoherence, maximum entanglement is obtained when the initial
qubit state is either spin-up or spin-down, while the equal superpostition state 
does not entangle during evolution (even though decoherence is observed). The choice 
of initial state also has a greater influence on the decay that follows, whereas
for pure decoherence, it only changes the amplitude of entanglement.

\section*{ACKNOWLEDGMENTS}
We thank Grzegorz Rajchel-Mieldzioć and Remigiusz Augusiak for the fruitful discussion on the upper bound for the negativity.
ICFO group acknowledges support from: ERC AdG NOQIA; Ministerio de Ciencia y Innovation Agencia Estatal de Investigaciones (PGC2018-097027-B-I00/10.13039/501100011033, CEX2019-000910-S/10.13039/501100011033, Plan National FIDEUA PID2019-106901GB-I00, FPI, QUANTERA MAQS PCI2019-111828-2, QUANTERA DYNAMITE PCI2022-132919, Proyectos de I+D+I “Retos Colaboración” QUSPIN RTC2019-007196-7); MCIN Recovery, Transformation and Resilience Plan with funding from European Union NextGenerationEU (PRTR C17.I1); Fundació Cellex; Fundació Mir-Puig; Generalitat de Catalunya (European Social Fund FEDER and CERCA program (AGAUR Grant No. 2017 SGR 134, QuantumCAT\~U16-011424, co-funded by ERDF Operational Program of Catalonia 2014-2020); Barcelona Supercomputing Center MareNostrum (FI-2022-1-0042); EU Horizon 2020 FET-OPEN OPTOlogic (Grant No 899794); National Science Centre, Poland (Symfonia Grant No. 2016/20/W/ST4/00314); European Union’s Horizon 2020 research and innovation programme under the Marie-Skłodowska-Curie grant agreement No 101029393 (STREDCH) and No 847648 (“La Caixa” Junior Leaders fellowships ID100010434: LCF/BQ/PI19/11690013, LCF/BQ/PI20/11760031, LCF/BQ/PR20/11770012, LCF/BQ/PR21/11840013).
M.P. acknowledges the support of the Polish National Agency for Academic Exchange, the Bekker programme no: PPN/BEK/2020/1/00317.

\bibliographystyle{apsrev4-2}
%


\end{document}